%% file: main.tex
\documentclass[sigconf]{acmart}

\usepackage{pgfplots, pgfplotstable}
\usepackage{url}
\usepackage{tikz}
\usepackage{natbib}
\usepackage{float}
\usepackage{etoolbox}
\usepackage{caption}
\usepackage{subcaption}
\usepackage{listings}
\usepackage{lstautogobble}
\usepackage{xcolor}
\usepackage{balance}
\usepackage{tikz}
\usepackage{fancyvrb}

\usepackage{draftwatermark}
\SetWatermarkText{Preprint}
\SetWatermarkScale{1}
\SetWatermarkLightness{0.9}

\definecolor{darksky}{rgb}{0.4,0.4,1}
\definecolor{skyblue}{rgb}{0.25,0.78,0.96}
\definecolor{lightyellow}{rgb}{1,0.96,0.52}
\definecolor{lightorange}{rgb}{1,0.7,0.4}
\definecolor{lightred}{rgb}{1,0.4,0.4}

\definecolor{codegreen}{rgb}{0,0.6,0}
\definecolor{codegray}{rgb}{0.5,0.5,0.5}
\definecolor{codepurple}{rgb}{0.58,0,0.82}
\definecolor{backcolour}{rgb}{0.95,0.95,0.92}

\lstdefinestyle{python}{
  backgroundcolor=\color{backcolour},
  commentstyle=\color{codegreen},
  keywordstyle=\color{magenta},
  numberstyle=\tiny\color{codegray},
  stringstyle=\color{codepurple},
  basicstyle=\ttfamily\footnotesize,
  breakatwhitespace=false,
  belowskip=-0.5em,
  breaklines=true,
  captionpos=b,
  keepspaces=true,
  numbersep=5pt,
  basicstyle=\footnotesize,
  showspaces=false,
  showstringspaces=false,
  showtabs=false,
  tabsize=4
}

\AtBeginDocument{%
  \providecommand\BibTeX{{%
\normalfont B\kern-0.5em{\scshape i\kern-0.25em b}\kern-0.8em\TeX}}}

\copyrightyear{2024}
\acmYear{2024}
\setcopyright{acmlicensed}\acmConference[ITiCSE 2024]{Proceedings of the 2024 Innovation and Technology in Computer Science Education V. 1}{July 8--10, 2024}{Milan, Italy}
\acmBooktitle{Proceedings of the 2024 Innovation and Technology in Computer Science Education V. 1 (ITiCSE 2024), July 8--10, 2024, Milan, Italy}
\acmDOI{10.1145/3649217.3653583}
\acmISBN{979-8-4007-0600-4/24/07}

\begin{document}

\title{Evaluating Micro Parsons Problems as Exam Questions}

\author{Zihan Wu}
\authornote{Both authors contributed equally to this research.}
\affiliation{%
   \institution{University of Michigan}
   \city{Ann Arbor}
   \state{MI}
 \country{USA}}
 \email{ziwu@umich.edu}
 \orcid{0000-0002-3161-2232}
 
 \author{David H. Smith IV}
 \authornotemark[1]
 \affiliation{%
   \institution{University of Illinois}
   \city{Urbana}
   \state{IL}
 \country{USA}}
 \email{dhsmith2@illinois.edu}
 \orcid{0000-0002-6572-4347}
 
\begin{abstract}
    Parsons problems are a type of programming activity that present learners with blocks of existing code and
    requiring them to arrange those blocks to form a program rather than write
    the code from scratch. 
    Micro Parsons problems extend this concept
    by having students assemble segments of code to form a single line of code rather
    than an entire program. Recent investigations into micro Parsons
    problems have primarily focused on supporting learners leaving open the
    question of micro Parsons efficacy as an exam item and how students perceive it when preparing for exams.

    To fill this gap, we included a variety of micro Parsons problems on four
    exams in an introductory programming course taught in Python. We use Item
    Response Theory to investigate the difficulty of the micro Parsons problems
    as well as the ability of the questions to differentiate between high and
    low ability students. We then compare these results to results for related
    questions where students are asked to write a single line of code from
    scratch. Finally, we conduct a thematic analysis of the survey responses to
    investigate how students' perceptions of micro Parsons both when practicing for exams
    and as they appear on exams.
\end{abstract}

\ccsdesc[500]{Social and professional topics~Computing education}

\keywords{Parsons Problems, CS1, Assessment, micro Parsons Problems}

\maketitle

\section{Introduction}\label{sec:intro}

\input{sections/intro.tex}

\section{Background}\label{sec:lit-review}

\input{sections/lit_review}

\section{Methods}\label{sec:methods}

\begin{figure*}
  \centering
  \begin{subfigure}[b]{0.49\textwidth}
    \includegraphics[width=\textwidth]{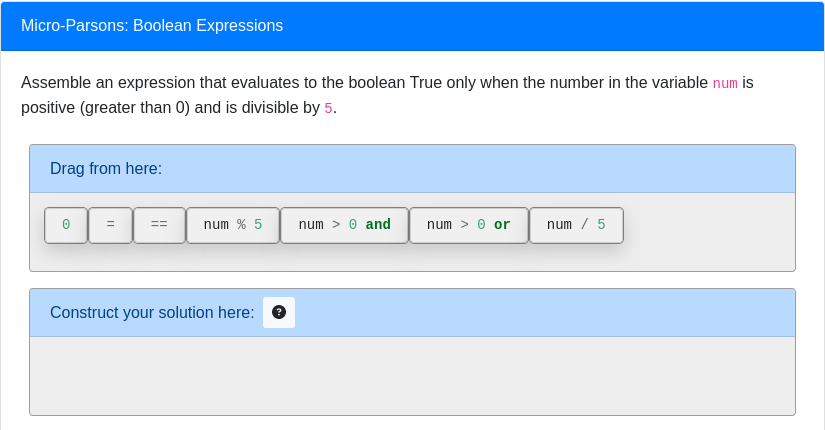}
    \caption{Standard micro Parsons - The first variation of micro Parsons problems as used on exams one and two.}
    \label{fig:micro-one}
  \end{subfigure}
  \hfill
  \begin{subfigure}[b]{0.49\textwidth}
    \includegraphics[width=\textwidth]{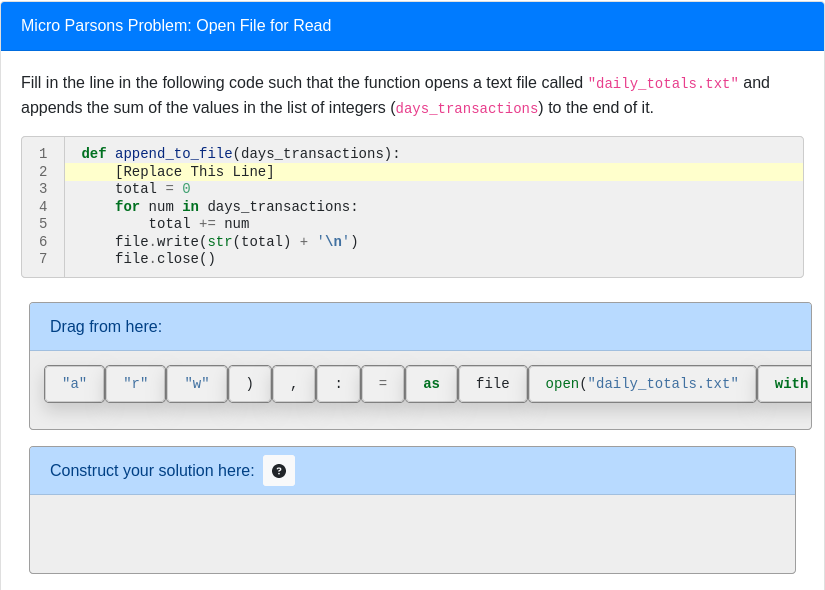}
    \caption{Fill-in-the-Blank micro Parsons - The second variation of Parsons problems as used on exams three and four.}
    \label{fig:micro-two}
  \end{subfigure}

  \caption{The two variations of Parsons problems used in the course.}
  \label{fig:micro-main}
\end{figure*}

\subsection{Course Context and Data Collection}

\begin{figure}[t]
    \centering
    \includegraphics[width=\columnwidth]{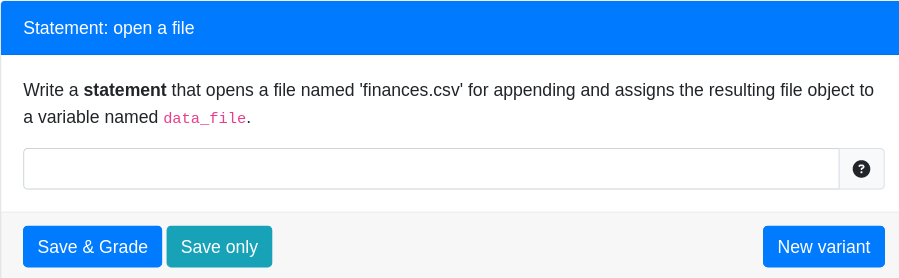}
    \caption{An example of a single-line programming question}
    \vspace{-5mm}
    \label{fig:slp}
\end{figure}

To evaluate our research questions, we designed a variety of micro Parsons
problems and deployed them on exams in a large introductory Python course. The
course aims at undergraduate students with non-tech majors such as business
majors, and covers basic Python topics including Python Collections, loops, and
basic classes and objects. Students are given four proctored exams throughout
the semester along with a practice exam generator released a week before each
exam. These exam generators allow students to continuously generate practice
exams, which draw questions from a bank of questions defined by the instructor.

The number of students sitting each exam varied throughout the semester due to
attrition, with 437 taking the first exam and 418 taking the final. All exams
are computer-based and delivered via an open-source assessment platform called
PrairieLearn~\cite{west2015prairielearn}. This platform enables students to
receive immediate feedback on the correctness of a submission and facilitates
partial credit by allowing students to make multiple attempts on a question for
a reduced number of points for each subsequent attempt. 

Beyond micro Parsons problems, these exams contain the following other question
types: 1) code writing question, 2) traditional parsons problems, 3) single-line
code writing problems (Figure~\ref{fig:slp}), 3) code fixing problems, 4) short
answer code comprehension questions, and 5) code tracing questions. Collection
of this data, as well as all other data collected in this course, was approved
by the institutional ethics review board at the institution where the study was
conducted.

\subsection{Problem Design}

Two variations of micro Parsons problems were deployed in the course. In the
first two exams the included problems asked students to arrange the given blocks to form a single,
independent line of code that accomplished a given task
(Figure~\ref{fig:micro-one}). In the remaining two exams, students were given an
existing segment of code with a single line of code removed. For these problems,
students were asked to arrange blocks to ``fill in the blank'' such that the
segment of code as a whole accomplished its given task (Figure~\ref{fig:micro-two}). In
both cases, the micro Parsons problems included jumbled distractors reduce the possibility of guessing the correct
solution. Additionally, students were given partial credit for incorrect attempts based on the number of blocks they positioned correctly.

\subsection{Student Survey}
In addressing \textbf{RQ2}, students were given an end-of-course survey on their
perceptions of the micro Parsons problems and how they compared to single-line
programming questions. First, students were asked to provide their general
thoughts in comparing the two question types in short answer format. Following
that, students were asked two multiple choice asking which question type they
preferred where the options were: (1) Micro Parsons, (2) Single Line Code
Writing, and (3) No Preference. The questions are as follows:
\begin{itemize}
    \item[\textbf{Q1:}] Which question type do you prefer to appear on exams?
    \item[\textbf{Q2:}] Which question type do you prefer to appear on study materials (e.g., practice exams)?
\end{itemize}
In association with each multiple choice question was a short answer question
which asked students to expand on their preference.

\begin{figure*}
  \centering
  \begin{subfigure}[b]{0.49\textwidth}
    \centering
    \includegraphics[width=\columnwidth]{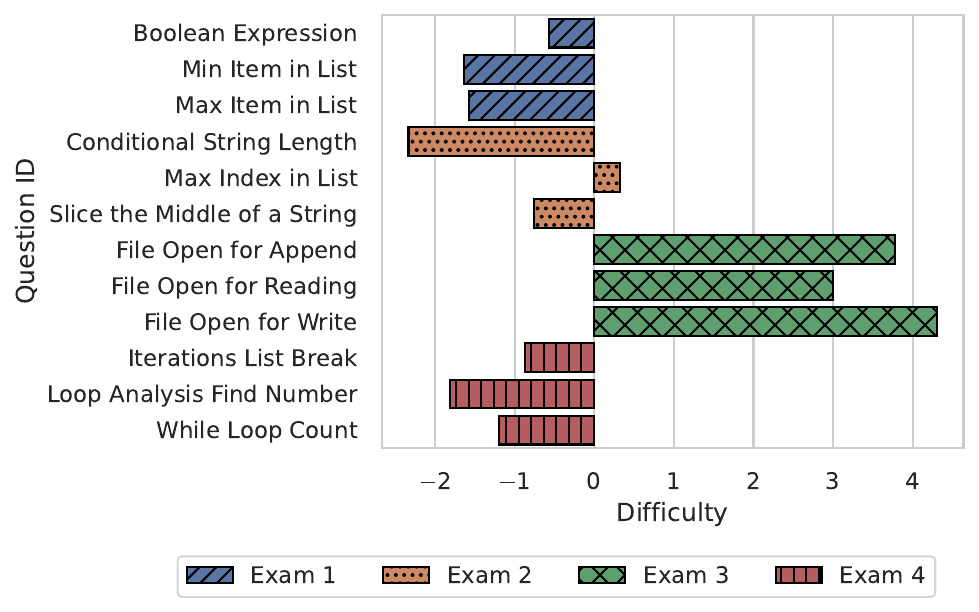}
    \vspace{-5mm}
    \caption{
        Item difficulty statistics from the 2PL model for each of the micro
        Parsons included on exams. Coefficients are interpreted as the ability
        level at which a student has a 50\% chance of answering 
        correctly.
    }
    \vspace{3.5mm}
    \label{fig:2pl-diff}
  \end{subfigure}
  \hfill
  \begin{subfigure}[b]{0.49\textwidth}
    \centering
    \includegraphics[width=\columnwidth]{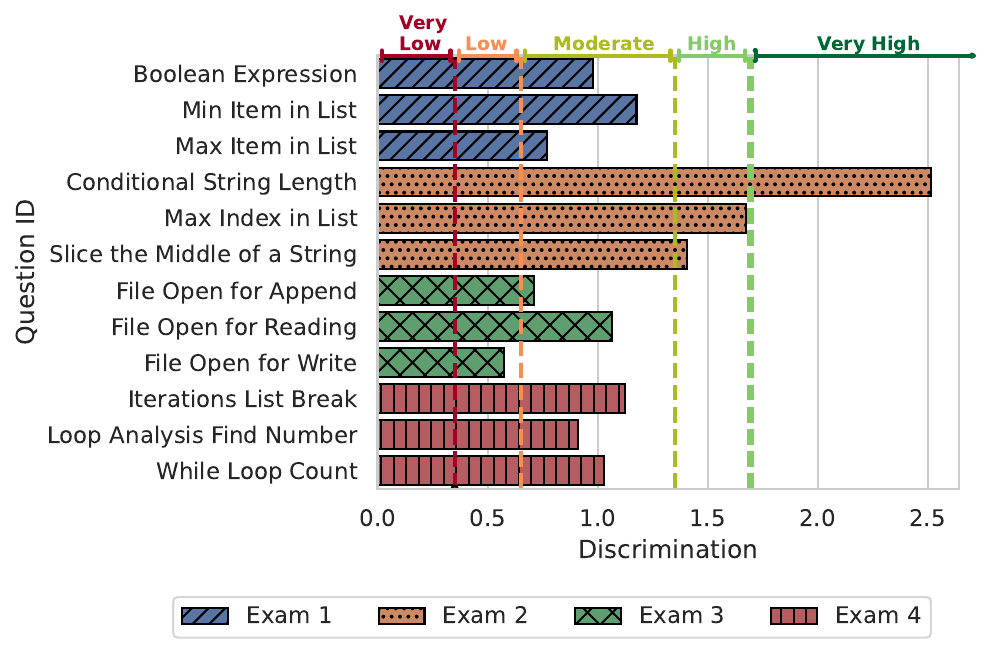}
    \vspace{-5mm}
    \caption{
    Item discrimination statistics from 2PL model for each of the micro Parsons
    included on exams. The thresholds for item discrimination (e.g., how well an
    item distinguishes between students above and below a given ability level)
    from \citet{baker2001basics} are presented as well. }
    \label{fig:2pl-discrim}
  \end{subfigure}

  \caption{The Item-Difficulty and Item-Discrimination statistics for the items used in this study.}
  \label{fig:statistics}
\end{figure*}

\begin{figure}[tb]
    \centering
    \includegraphics[width=0.95\columnwidth]{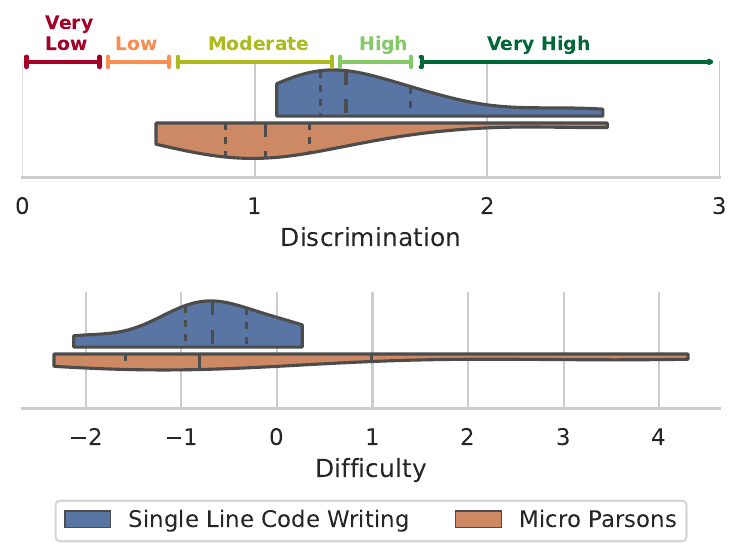}
    \vspace{-2mm}
    \caption{
    Comparison of the distributions of item-difficulty and item-discrimination
    statistics for micro Parsons and single-line code writing questions
    }
    \label{fig:discrim_diff_comparision}
    \vspace{-3mm}
\end{figure}

\section{RQ1: Difficulty and Discrimination}
\label{sec:rq1}
To address \textbf{RQ1} the scores students received on questions were
dichotomized based on the success of a student's first attempt. This is done to
enable analysis with a 2PL IRT model, which requires the use of dichotomized
scores. The reason for fitting the model based on the correctness
of a student's first attempt rather than their best attempt is, given students
can attempt a question multiple times, many questions were
ultimately answered correctly by all or a large majority of students. This would
prevent those questions from being included when fitting the model, as 2PL IRT does not support the
inclusion of questions with one response category (i.e., all correct or all
incorrect). We fit a 2PL IRT model for each exam to determine the
 item difficulty and item discrimination statistics for the items that appear on
those exams. To aid in the interpretation of the item-discrimination statistics,
figures use the thresholds presented by \citet{baker2001basics} of: Very Low
($\leq0.34$), Low ($0.35-0.64$), Moderate ($0.65-1.34$), High ($1.35-1.69$), and Very High ($>1.7$).

For the comparison between single-line code writing problems and micro Parsons problems, we selected a total of 17 single-line code writing questions that appeared alongside the micro Parsons problems on exams. These questions cover related topics (e.g., string slicing, boolean expressions, file opening) but are not identical in format to the micro Parsons. As such, their results are included to help contextualize the item statistics for micro Parsons -- given they are a related item testing similar concepts -- rather than provide a one-to-one comparison.

\subsection{Results}
\label{sec:stats-results}
Overall, the micro Parsons problems appeared to generally have low difficulty
(Figure~\ref{fig:2pl-diff}) coefficients with moderate discrimination
(Figure~\ref{fig:2pl-discrim}). These results indicate that the problems, in
general, appear to do a reasonable job of differentiating between students is
the lower ability range. Comparing the distribution of difficulty and
discrimination statistics for micro Parsons and single-line programming
questions we find that both appear to have similar distributions of difficulty
but single-line code writing problems have higher item-discrimination
(Figure~\ref{fig:discrim_diff_comparision}). This indicates that on average the
item functions for students of lower ability and single-line code writing questions provide a
more sharp delineation between students at that ability level. 

In looking at Figure~\ref{fig:2pl-diff}, the one notable exception to this
trend are the questions relating to opening a file which appeared on exam three.
These questions had exceptionally high difficulty compared to micro Parsons
which appeared on other exams. These questions differed from their counterparts in
that the student was required to read the surrounding code and, based on the
indentation, determine whether to open the file using only the \texttt{open()}
function or using \texttt{with-as}. In looking at the incorrect submissions we
find that 34\% of all incorrect submissions attempted to use \texttt{with-as}.
This highlights a potential connection exists between the distractors that are
chosen in micro Parsons and the provided code for which the students are
intended to fill in the blanks.

\section{RQ2: Student Perceptions and Preferences}
\label{sec:rq2}
In total, we collected survey responses from 390 students.
Two researchers independently coded the first 50 responses to develop an initial code book.
Upon iterating and finalizing the code book on the first 50 responses, the researchers independently coded the next 50 to establish an interrater agreement.
The inter-rater agreement, Krippendorf's alpha, was $\alpha = 0.7$, which was sufficient for preliminary conclusion~\cite{krippendorff2018content}.
One of the researchers coded the remaining 290 responses according to the established code book.

\subsection{Results}

We collected anonymous responses from 390 students asking which question types they preferred, both in the context of exams and practice.
When asked about the preferred type of question to appear on the exam, 163 (41\%) students chose micro Parsons problems, 138 (35\%) chose single-line code writing questions, and 89 (23\%) students chose no preference.
For practice, however, the most popular response was no preference (n = 160, 41\%), while 134 (34\%) students preferred single-line code writing questions, and 96 (25\%) students preferred micro Parsons problems.

Several themes emerged from students’ responses in explaining their choices and comparing the two types of problems.

\subsubsection{Perceived Difficulty}
When comparing the two types of problems, 83 students (21\%) explicitly noted that they thought micro Parsons problems were easier.
Among them, 13 students felt that micro Parsons problems \textbf{provided hints} compared to the single-line code writing question, especially when they did not know where to start: \textit{“micro parsons are good to give you a head start on the topic and help you out a bit, [single-line code writing] questions kind of mess you up if you don't know where to start.” (P308)}
Fifteen students felt that micro Parsons problems were easier in exams because they provided \textbf{partial credit} for incorrect responses.

Contrary to prior within-subjects think-aloud studies on Parsons problems where all participants find them easier than writing code from scratch~\cite{haynes2021problem}, in our survey responses, 50 students (13\%) felt that micro Parsons problems are more difficult than single-line code writing problems.
\textit{“I found that the micro Parsons were slightly more difficult as I had to try to think through the code with what was given instead of trying to think it through on my own.” (P23)}
In our study, only micro Parsons problems contained the fill-in-the-blank type of problems, which can also contribute to the perceived difficulty.

\vspace{-0.5mm} 

\subsubsection{Assisting Learning}
Most students reported that they found both types of problems beneficial for learning.
Meanwhile, 36 students (9\%) specifically pointed out that micro Parsons are helpful for learning, because they \textbf{struggle less} and \textbf{can practice specific parts of a single line}. They felt that micro Parsons problems  \textbf{provide examples} of how the code lines should look like, \textbf{provide better feedback} on which part of the solution was incorrect, and \textbf{require less memorization}: “\textit{it is more in line with how actual coding is: you do not have to memorize a lot of stuff, you should know the concept and the ways to look for the information.}” (P98)
On the other hand, 32 (8\%) students explicitly mentioned that the single-line code-writing questions helped them learn. 
They felt these questions involve \textbf{“actual coding”} and \textbf{shows what they actually know}.
They felt that having less help could force them \textbf{memorize better} or apply more \textbf{high-level thinking}.

Interestingly, learners’ comments on which variation they think better helps learning do not always align with their selected preferences for homework or exams.
Many students explicitly said they preferred the practice exams to \textit{“have the same format of questions” that would appear on the exam. (P159)}
For the exam questions, students chose micro Parsons problems because they provide extra credits when the answers are incorrect.

\subsubsection{Confusing Distractors}
A total of 40 students (10\%) mentioned that micro Parsons problems could be confusing at times.
Specifically, 10 students offered complaints about the distractors in the micro Parsons problems. Some suggested removing all distractors, or providing a “use all blocks” prompt for those problems that do not contain distractors:
\textit{“Usually with the micro Parsons problems I tend to overthink my answer and end up getting the question wrong. All the extra blocks are confusing.” (P215)}
Most of the time, the distractors contain minor problems that novices tend to make, such as the use of square brackets instead of parentheses, or ignoring quotation marks around string literals, and minor logic errors (e.g., \texttt{and} vs \texttt{or}).
Four students explicitly expressed their annoyance with these distractors:
\textit{“it is extremely annoying when micro parsons problems include wrong blocks because it only feels like they are trying to trip me up based on reading errors.” (P292)}

\section{Discussion}
\label{sec:discuss}

From our results we highlight and expand on three takeaways: 1) the ability of
micro Parsons problems to provide partial credit and simplified feedback; 2)
commentary on how distractor selection can in some cases significantly impact
item difficulty; 3) the expertise reversal effect and its potential impact
on the design of micro Parsons; and 4) the relationship between the presence of distractors and confusion students encountered on micro Parsons problems. Each
of these serves to inform the design and use of micro Parsons as an item used on
exams and practice materials alike.

\subsection{Feedback and Partial Credit with Micro Parsons Problems}
\begin{figure}[tb]
    \centering
    \includegraphics[width=\columnwidth]{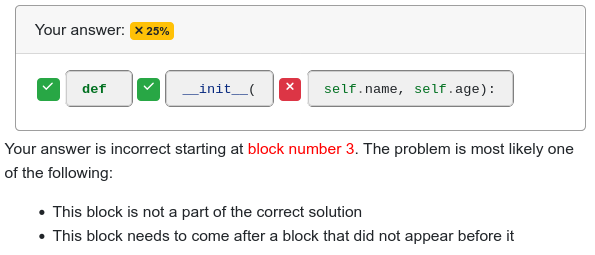}
    \vspace{-6mm}
    \caption{
        The feedback and partial credit mechanism used for our implementation of
        micro Parsons. 
    }
    \vspace{-3mm}
    \label{fig:enter-label}
\end{figure}

As noted by students in the qualitative results, one of the key differences
between micro Parsons problems and single-line code writing problems is the
limited solution space enforced by requiring students to arrange predefined
blocks. As noted by \citet{denny2008evaluating} -- though in the context of
traditional Parsons problems -- this enables more straightforward scoring for instructors. However, a notable difference between the
micro Parsons used in this study and those used in the investigations of
\citet{denny2008evaluating} is our study uses an online platform to deliver the
questions whereas their used a paper based exam. The ability to receive instantaneous feedback and make multiple attempts allows forwards these benefits onto students as well both during the exam and while practicing.

As noted by many students, the use of single-line code writing problems and micro Parsons alike was well
received as many students viewed as questions as testing them on and allowing
them to practice elements of larger programs in isolation. Though it may be the
case that micro Parsons problems have less discriminating power relative to
single-line code writing questions there are other considerations that
instructors may have beyond these metrics. By reducing the error space and
providing feedback that is easier to interpret for this class of programming
problem it is likely that the students' experience is improved. This is
particularly a consideration in the context of practice exams, such as the ones
our students encountered, as it scaffolds their studying process.

\subsection{Distractor Selection for Fill-in-the-Blank}

As discussed in the quantitative results (Section~\ref{sec:rq1}), the
fill-in-the-blank micro Parsons (see fig.~\ref{fig:micro-two}) had much higher item-difficulties any other
micro Parsons problem or single line code writing question. In these questions,
students were presented with all the blocks needed to open a file either using
\texttt{with-as} or simply storing the result of \texttt{open()} in a variable.
The intention was for students to infer from the code they were completing that 
\texttt{with-as} was not appropriate given it requires any manipulation of the file
to be done underneath it in an indented block. This problem highlights a unique
connection that can be made between the distractors chosen and the context of
the code the student is required to complete in micro Parsons problems for
languages where whitespace is semantic.

\subsection{Potential Expertise Reversal Effect}

From the student survey, we discovered that some
students perceived micro Parsons problems as more difficult than
writing code from scratch. This finding differs from prior work that
included within-subject studies for students to compare micro Parsons problems
or regular Parsons problems with writing code from scratch~\cite{wu2023using,
haynes2021problem}. One potential explanation is the expertise reversal
effect. It refers to the situation where some
instructional methods designed for novices become inefficient or have
negative consequences for advanced learners~\cite{kalyuga2009expertise}.
According to cognitive load theory, the instructional methods for novices
usually contain extra information and guidance. However, expert learners already
possess schema-based knowledge, which provides internal guidance. Expert
learners need to resolve and reconcile the two different sets of guidance, which
can introduce extra burdens in working memory.

As pointed out by some students in the qualitative analysis, constructing
solutions only by given blocks can sometimes be even more difficult when they
already have a plain-code answer in mind. The hints, in this case, the blocks,
can become extra guidance that is not needed by expert learners and cause a
burden for them. This is particularly a consideration when selecting distractor
blocks as the inclusion of plausible alternatives -- and thus more blocks -- 
may increase the prevalence and impact of this effect. This may be
considered a desirable difficulty in practice materials for illustrating
alternative solutions. However, in summative assessments this added difficulty
may stem from the question format rather than the content being tested. This finding
highlights the careful consideration that should be taken when selecting the
number of blocks, and distractors, that are included in a solution space.

\subsection{Confusing Distractors}

In our qualitative results, we discovered that some students found micro Parsons
problems confusing. The major complaint is the distractors. Only a few students
elaborated on their confusion, explaining that they felt that the distractors
were too tricky. There are two interpretations behind students' perception of
the trickiness of the distractors. First, students felt that sometimes it was
too difficult to identify the correct blocks versus the distractors; It is also
possible that the complaints reflected that they deem the difficulty
unnecessary.

Although distractors received complaints from some students, prior work in the
context of traditional Parsons problems suggests they may be useful for
learning. Prior work on proof blocks, a type of Parsons problem for mathematical
proofs has shown that learners who completed proof blocks with distractors
performed better on the post-test than those who completed proof blocks without
distractors, though the difference was not statistically
significant~\cite{poulsen2023measuring}. In a multi-institutional multi-national
study, learners who practiced with distractors were also found to have better
performances on fixing code that contains similar errors in the distractors than
those who did not~\cite{ericson2023multi}. Considering the importance of
debugging and fixing code, especially with the development of generative AI
tools, it is likely distractors will become increasingly important for learning.

Meanwhile, the visual presentation of the distractors can also affect students'
perceptions. Similar to previous research on micro Parsons problems, we only
used jumbled distractors in this study. Because micro Parsons problems require
learners to rearrange code in one line, the design of interfaces for visual
grouping distractors with their correct alternatives is not as straightforward.
In our study, no student provided concrete examples of distractors that they
found confusing. Future research should investigate the attributes that make
distractors confusing for learners, and examine the effective ways to add
distractors that can reduce confusion while maintaining item difficulty and item
discrimination.

\section{Limitations and Future Work}\label{sec:limits}

There are several limitations to the results of our study that can affect its
generalizability. First, only a small number of micro Parsons were used
throughout the semester. More questions and question versions would need to be
created to provide a more complete and accurate picture of how micro Parsons
function as an exam item and how they compare to related text-entry code writing
questions.

Second, a limitation of the quantitative analysis of this work is that the
comparison of item-discrimination statistics between micro Parsons problems and
single-line code writing problems was at a categorical level. That is, we did
not create pairs of questions with identical solutions and compare them. As
such, our results comparing the item discrimination and difficulty distributions
are more suited to contextualize the results for micro Parsons using a related
code writing exercise. 

A final limitation is students had access to practice exams which contained both
the single line code writing problem generators used on exams and micro Parsons
that covered related topics to those on their exams. This likely had the impact
of lowering the difficulty of the items compared to if fully novel questions
were used on the exams.

\section{Conclusion}\label{sec:conclude}
This study employed micro Parsons problems as exam items across four exams, and
adopted measurement theory to analyze its effectiveness. We calculated the item
difficulty and item discrimination of micro Parsons problems as well as
traditional single-line code writing problems across the semester. In this study, we
found that micro Parsons questions are comparable in difficulty but slightly
lower in discrimination compared to writing code from scratch. We also found
insights that are worth future exploration from our student survey. Students
displayed diverse preferences when it came to micro Parsons problems and
single-line code questions in exam, largely based on their perceived potential
to earn more points. We found a potential expertise reversal effect from student
response that considered micro Parsons problems harder than writing code from
scratch, which motivates further inquiries into the ideal design for micro
Parsons problems used in exams.

\bibliographystyle{ACM-Reference-Format}
\balance
\bibliography{acmart}

\end{document}

%% file: sections/intro.tex
Learning to program is difficult for beginners.
Researchers have been working to design pedagogical activities for learners that are engaging and welcoming for novices.
To improve student engagement, present models for good practice, and provide immediate feedback, Parsons and Haden~\cite{parsons2006parson} designed Parsons problems as an alternative type of practice.
Instead of asking learners to write code from scratch, Parsons problems provide mixed-up code blocks and require learners to identify the blocks needed in the correct solution and rearrange them into the correct order.
As a type of completion problem, Parsons problems make programming practice less challenging for beginners by reducing the problem space.

In traditional Parsons problems, each code block contains at least one line of code.
Micro Parsons problems extended this concept by focusing on a single line of code.
It provides small code fragments for learners and asks them to form a code statement in one line.
Prior work on micro Parsons problems has studied its effect on learners as practice questions, and found that it reduced the time needed for learners to complete the problems, encouraged more learners to complete optional practice tasks, and resulted in comparable learning gain as writing code from scratch~\cite{wu2023using}.

For traditional Parsons problems, \citet{denny2008evaluating} evaluated their effectiveness as exam items, and found that students' performances on these problems are highly correlated with code writing tasks.
As a fine-grained variation of Parsons problems, micro Parsons problems also have the potential to be used as exam items.
However, existing literature on micro Parsons problems focuses on their benefits for learners, leaving an unaddressed question concerning their effectiveness as exam items. 

This paper aims to bridge this gap by conducting an empirical investigation of using micro Parsons problems within the context of exams and exam preparation materials in an introductory programming course.
We investigate the following research questions:
\begin{enumerate}
    \item[\textbf{RQ1:}] What are the psychometric properties (e.g., item difficulty and discrimination) of micro Parsons problems and how do these compare to those of single line code writing problems?
    \item[\textbf{RQ2:}] What are students perceptions and preferences when comparing micro Parsons problems to single line code writing problems for the purposes of studying and when they appear on summative assessments?
\end{enumerate}

%% file: sections/lit_review.tex
\subsection{Parsons and Micro Parsons Problems}

Parsons problems, as introduced by \citet{parsons2006parson}, are a programming
activity typically used in introductory courses.  They provide a scaffolded
environment to construct programs, presenting students with blocks of code that
they must rearrange to form a solution rather than having them write that code
from scratch. This problem type has shown a variety of benefits for learners
such as improving learning efficiency~\cite{ericson2018evaluating}, improving
engagement~\cite{parsons2006parson, morin2020work, ericson2015analysis},
reducing cognitive load~\cite{bender2023integrating, fabic2019evaluation,
morrison2016subgoals}, improving the motivation of
students~\cite{kumar2017effect}. Researchers have designed many variants of
Parsons problems and investigated students'
perceptions~\cite{morin2021collaborative, oyelere2019impact,
denny2008evaluating}.

A recently introduced variant of Parsons problems is micro Parsons
problems~\cite{wu2021regex, wu2023investigating}. Whereas in traditional Parsons
problems students are asked to vertically arrange blocks that each contain at
least one line of code, micro Parsons problems require students to arrange
smaller code fragments horizontally to form a single line of code. The
motivation for this variation is to transfer the positive impacts that have been
found with respect to traditional Parsons problems to other domains where the
construction of a single line of code or expression is the core learning
objective (e.g., regex, SQL). Much like prior investigations into traditional
Parsons problems, initial investigations into micro Parsons have also focused on
the impacts of utilizing them as a tool for learning~\cite{wu2023using}. Given
the recency of this variety of Parsons problem, additional investigations to
fully explore the effectiveness of this problem type in both formative and
summative contexts are warranted.

\subsection{Parsons Problems as Exam Items}

Though both the original purpose of Parsons problems and much of the research
that has followed has focused on their use as a tool for
learning~\cite{ericson2022parsons}, several studies have investigated their
utility as an exam item. \citet{denny2008evaluating} provided the first of these
investigations. They identified that students' performance in Parsons problems
was highly correlated with performance on code writing tasks, suggesting both
items measure a similar skill. Similarly, \citet{poulsen2022evaluating} found
that proof writing questions presented in a format similar to Parsons Problems,
termed ``Proof Blocks'', still provided ample information on students' proof
writing abilities. 

\citet{denny2008evaluating} make additional recommendations
on the design of Parsons problems for exams suggesting that: (1) incorrect
blocks of code (termed distractors) should be visually grouped with their
correct alternatives to minimize cognitive load and (2) the inclusion of syntax
that indicates structure (e.g., curly braces, indentation) can be used to
provide hints on the correct response. With respect to the former design
consideration involving the use of distractors, more recent work has indicated
that the presence of distractors increasing the difficulty of the item compared
to not having distractors. However, their presence was shown to have a minimal
impact on Parsons problems quality as an exam item at the cost of students
spending substantially more time on the problem~\cite{smith2023discovering,
smith2023investigating, smith2023useful, smith2023comparing}. Similar
investigations informing the design and utility of micro Parsons as an exam item
have yet to be performed.

\subsection{Measurement Theory}

Central to evaluating a question type as an exam item is selecting a tool set
for performing that evaluation. Additionally, the goal of the assessment must be
identified and in order to determine how its items can be improved to meet that
goal. There are multiple ways to evaluate the type of questions. In the case of
the study by \citet{denny2008evaluating}, one of the goals was to identify if
Parsons problems evaluated a skill similar to code writing. Other criteria
outside of this work include the difficulty of the item, and the ability of the
question to delineate between students who are adept at a given skill and those
who are not. By improving these characteristics of the items that make up an
exam we can improve the measurement ability of the exam as a
whole~\cite{ebel1972essentials}.

A common approach for examining the characteristics of an item is Item Response
Theory (IRT). IRT models the probability of a student of a given ability
($\theta$) to respond to a problem with a variety of estimated parameters that
characterize a given item. There are a variety of models in the IRT family, each
characterized by the number of parameters it includes. For example, the 2
Parameter Logistic Model (2PL) \[P_i(\theta) = \frac{1}{1 + e^{-(a_i \cdot
(\theta - b_i))}}\] includes parameters for estimating item discrimination ($a$)
and difficulty ($b$)~\cite{de2010primer, birnbaum1968some}. Difficulty is
characterized by the ability level at which a student has a 50\% chance of
responding correctly to a question and discrimination how sharply an item
distinguishes between students at a that given difficulty level. The 3PL model
adds a third parameter generally referred to as ``pseudo-guessing'' which
estimates a lower asymptotic bound and is interpreted as the probability of a
low-ability student guessing the correct solution~\cite{birnbaum1968some}. The
4PL model adds a final fourth parameter for estimating an upper asymptotic bound
which is often referred to as ``carelessness'' or ``slip'' and is interpreted as
the probability of an otherwise high ability student responding incorrectly to
the question~\cite{barton1981upper}.